\documentclass[sigplan, 10pt]{acmart}

\renewcommand\footnotetextcopyrightpermission[1]{}
\AtBeginDocument{%
  \providecommand\BibTeX{{%
    \normalfont B\kern-0.5em{\scshape i\kern-0.25em b}\kern-0.8em\TeX}}}

\usepackage{tikz}
\usepackage{amsmath}

\usepackage{filecontents}

\usepackage{times}
\usepackage{datetime}
\usepackage{url}
\usepackage{hyperref}

\usepackage[normalem]{ulem}
\usepackage{multirow}
\usepackage{comment}
\usepackage{listings}
\usepackage{tikz}
\usepackage{verbatimbox}

\usepackage{amssymb}
\usepackage{pifont}
\usepackage{booktabs}
\usepackage[keeplastbox]{flushend}
\usepackage{enumitem}
\usepackage{caption}

\newcommand{\todo}[1]{{\color{red}\bfseries [[#1]]}}

\newcommand{\NP}{\texttt{NumaPerf}}
\newcommand{\specialcell}[2][c]{%
  \begin{tabular}[#1]{@{}c@{}}#2\end{tabular}}
  \newcommand{\TM}{\text{thread migration}}
  \newcommand{\PS}{\text{remote access}}
  \newcommand{\TS}{\text{true sharing}}
  \newcommand{\FS}{\text{false sharing}}
  \newcommand{\TI}{\text{thread imbalance}}   \newcommand{\TB}{\text{thread binding}}
  \newcommand{\BI}{\text{block interleave}}
  \newcommand{\PI}{\text{page interleave}}
  \newcommand{\PAD}{\text{padding}}
  \newcommand{\DUP}{\text{duplicate}}

\begin{document}

\date{}

\title{NumaPerf: Predictive and Full NUMA Profiling}



\author{
Xin Zhao, Jin Zhou, and Hui Guan, University of Massachusetts Amherst }
\author{
Wei Wang, University of Texas at San Antonio}
\author{
Xu Liu, North Carolina State University }
\author{
Tongping Liu, University of Massachusetts Amherst
} 


\begin{abstract}
Parallel applications are extremely challenging to achieve the optimal performance on the NUMA architecture,  which necessitates the assistance of profiling tools. However, existing NUMA-profiling tools share some similar shortcomings, such as portability, effectiveness, and helpfulness issues. This paper proposes a novel profiling tool--\NP{}--that overcomes these issues.  \NP{} aims to identify potential performance issues for any NUMA architecture, instead of only on the current hardware. To achieve this, \NP{} focuses on memory sharing patterns between threads, instead of real remote accesses. \NP{} further detects potential thread migrations and load imbalance issues that could significantly affect the performance but are omitted by existing profilers. \NP{} also separates cache coherence issues that may require different fix strategies. Based on our extensive evaluation, \NP{} is able to identify more performance issues than any existing tool, while fixing these bugs leads to up to $5.94\times$ performance speedup. 

\end{abstract} 

\maketitle

\section{Introduction}
\label{sec:intro}

The Non-Uniform Memory Access (NUMA) is the de facto design to address the scalability issue with an increased number of hardware cores. Compared to the Uniform Memory Access (UMA) architecture, the NUMA architecture avoids the bottleneck of one memory controller by allowing each node/processor to concurrently access its own memory controller. However, the NUMA architecture imposes multiple system challenges for writing efficient parallel applications, such as remote accesses, interconnect congestion, and node imbalance~\cite{Blagodurov:2011:CNC:2002181.2002182}. User programs could easily suffer from significant performance degradation, necessitating the development of profiling tools to identify NUMA-related performance issues. 

General-purpose profilers, such as \texttt{gprof}~\cite{DBLP:conf/sigplan/GrahamKM82}, \texttt{perf}~\cite{perf}, or \texttt{Coz}~\cite{Coz}, are not suitable for identifying NUMA-related performance issues~\cite{XuNuma,valat:2018:numaprof} because they are agnostic to the architecture difference. 
To detect NUMA-related issues, one type of tools
simulates cache activities and page affinity based on the collected memory traces~\cite{NUMAGrind, MACPO}. However, they may introduce significant performance slowdown, preventing their uses even in development phases. In addition to this, another type of profilers employs coarse-grained sampling to identify performance issues in the deployment environment~\cite{Intel:VTune, Memphis, Lachaize:2012:MMP:2342821.2342826, XuNuma, NumaMMA, 7847070}, while the third type builds on fine-grained instrumentation that could detect more performance issues but with a higher overhead~\cite{diener2015characterizing, valat:2018:numaprof}. 


However, the latter two types of tools share the following \textbf{common issues}. \textit{First, they mainly focus on one type of performance issues (i.e., remote accesses), while omitting other types of issues that may have a larger performance impact}. 
\textit{Second, they have limited portability that can only identify remote accesses on the current NUMA hardware}. The major reason is that they rely on the physical node information to detect remote accesses, where the physical page a thread accesses is located in a node that is different from the node of the current thread. However, 
the relationship between threads/pages with physical nodes can be varied when an application is running on different hardware with different topology, or even on the same hardware at another time. That is, existing tools may miss some remote accesses caused by specific binding. \textit{Third, existing tools could not provide sufficient guidelines for bug fixes}. Users have to spend significant effort to figure out the corresponding fix strategy by themselves.



This paper proposes a novel tool---\NP{}---that overcomes these issues. \NP{} is designed as an automatic tool that does not require human annotation or the change of the code. It also does not require new hardware, or the change of the underlying operating system. \NP{} aims to detect NUMA-related issues in development phases, when applications are exercised with representative inputs. In this way, there is no need to pay additional and unnecessary runtime overhead in deployment phases. We further describe \NP{}'s distinctive goals and designs as follows. 

First, \NP{} aims to detect some additional types of NUMA performance issues, while existing NUMA profilers could only detect remote access. The first type is load imbalance among threads, which may lead to memory controller congestion and interconnect congestion. The second type is cross-node migration, which turns all previous local accesses into remote accesses. Based on our evaluation, cross-node migration may lead to $4.2\times$ performance degradation for \texttt{fluidanimate}. However, some applications may not have such issues, which requires the assistance of profiling tools.  



Second, it proposes a set of architecture-independent and scheduling-independent mechanisms that could predictively detect the above-mentioned issues on any NUMA architecture, even without running on a NUMA machine. \NP{}'s detection of remote accesses is based on a \textbf{key observation}:  memory sharing pattern of threads is an invariant determined by the program logic, but the relationship between threads/pages and physical nodes is architecture and scheduling dependent. Therefore, \textit{\NP{} focuses on identifying memory sharing pattern between threads, instead of the specific node relationship of threads and pages}, since a thread/page can be scheduled/allocated to/from a different node in a different execution. This mechanism not only simplifies the detection problem (without the need to track the node information), but also generalizes to different architectures and executions (scheduling). \textit{\NP{} also proposes an architecture-independent mechanism to measure load imbalance based on the total number of memory accesses from threads}: when different types of threads have a different number of total memory accesses, then this application has a load imbalance issue. \textit{\NP{} further proposes a method to predict the probability of thread migrations.} \NP{} computes a migration score based on the contending number of synchronizations, and the number of condition and barrier waits. Overall, \NP{} predicts a set of NUMA performance issues without the requirement of testing on a NUMA machine, where its basic ideas are further discussed in Section~\ref{sec:idea}.   




Last but not least, \NP{} aims to provide more helpful information to assist bug fixes. 
Firstly, it proposes a set of metrics to measure the seriousness of different performance issues, preventing programmers from spending unnecessary efforts on insignificant issues. Secondly, its report could guide users for a better fix. For load imbalance issues, \NP{} suggests a thread assignment that could achieve much better performance than existing work~\cite{SyncPerf}. For remote accesses, there exist multiple fix strategies with different levels of improvement. Currently, programmers have to figure out a good strategy by themselves. In contrast, \NP{} supplies more information to assist fixes. It separates cache false sharing issues from true sharing and page sharing so that users can use the padding to achieve better performance. It further reports whether the data can be duplicated or not by confirming the temporal relationship of memory reads/writes. It also reports threads accessing each page, which helps confirm whether a block-wise interleave with the thread binding will have a better performance improvement. 



We performed extensive experiments to verify the effectiveness of \NP{} with widely-used parallel applications (i.e., PARSEC~\cite{parsec}) and HPC applications (e.g., AMG2006~\cite{AMG2006}, Lulesh~\cite{LULESH}, and UMT2003~\cite{UMT2013}).  Based on our evaluation, \NP{} detects many more performance issues than the combination of all existing NUMA profilers, including both fine-grained and coarse-grained tools. After fixing such issues, these applications could achieve up to $5.94\times$ performance improvement.  \NP{}'s helpfulness on bug fixes is also exemplified by multiple case studies. Overall, \NP{} imposes less than $6\times$ performance overhead, which is orders of magnitude faster than the previous state-of-the-art in the fine-grained analysis. The experiments also confirm that \NP{}'s detection is architecture-independent, which is able to identify most performance issues when running on a non-NUMA machine. 

Overall, \NP{} makes the following contributions. 

\begin{itemize}
    \item \NP{} proposes a set of architecture-independent and scheduling-independent methods that could predictively detect NUMA-related performance issues, even without evaluating on a specific NUMA architecture. 
    \item \NP{} is able to detect a comprehensive set of NUMA-related performance issues, where some are omitted by existing tools. 
    
    \item \NP{} designs a set of metrics to measure the seriousness of performance issues, and provides helpful information to assist bug fixes.
    \item We have performed extensive evaluations to confirm \NP{}'s effectiveness and overhead.  
\end{itemize}

\subsection*{Outline}

The remainder of this paper is organized as follows. Section~\ref{sec:overview} introduces the background of NUMA architecture and the basic ideas of \NP{}. Then Section~\ref{sec:implementation} presents the detailed implementation and Section~\ref{sec:evaluation} shows experimental results. After that, Section~\ref{sec:discussion} explains the limitation and  Section~\ref{sec:related} discusses related work in this field. In the end, Section~\ref{sec:conclusion} concludes this paper.

\section{Background and Overview}
\label{sec:overview}

This section starts with the introduction of the NUMA architecture and potential performance issues. Then it briefly discusses the basic idea of \NP{} to identify such issues. 

\subsection{NUMA Architecture}
\label{sec:numa}

Traditional computers use the Uniform Memory Access (UMA) model. In this model,  all CPU cores share a single memory controller such that any core can access the memory with the same latency (uniformly). However, the UMA architecture cannot accommodate the increasing number of cores because these cores may compete for the same memory controller. The memory controller becomes the performance bottleneck in many-core machines since a task cannot proceed without getting its necessary data from the memory. 

\begin{figure}[htbp]
\centering
\includegraphics[width=0.9\columnwidth]{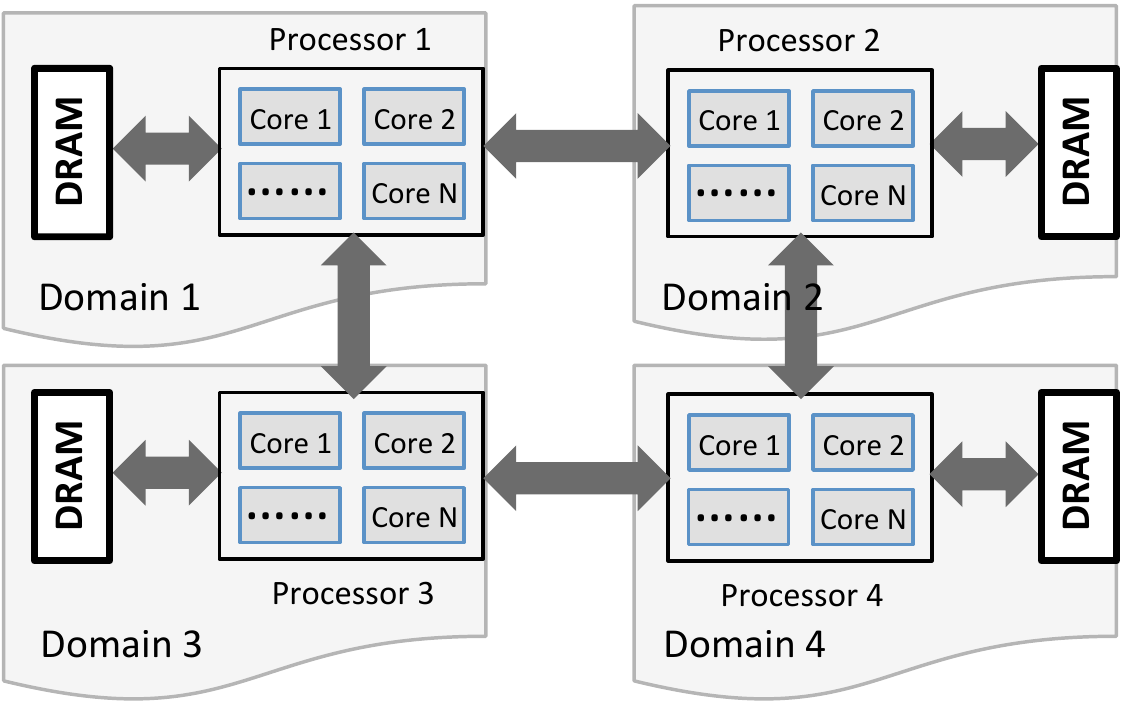}
\caption{A NUMA architecture with four nodes/domains\label{fig:numa}}
\end{figure}

The Non-Uniform Memory Access (NUMA) architecture is proposed to solve this scalability issue, as further shown in Figure~\ref{fig:numa}. It has a decentralized nature. Instead of making all cores waiting for the same memory controller, the NUMA architecture is typically equipped with multiple memory controllers, where each controller serves a group of CPU cores (called a ``node'' or ``processor'' interchangeably). Incorporating multiple memory controllers largely reduces the contention for memory controllers and therefore improves the scalability correspondingly. However, the NUMA architecture also introduce multiple sources of performance degradations~\cite{Blagodurov:2011:CNC:2002181.2002182}, including \textit{Cache Contention}, \textit{Node Imbalance}, \textit{Interconnect Congestion}, and \textit{Remote Accesses}. 

\textbf{Cache Contention:} the NUMA architecture is prone to cache contention, including false and true sharing. False sharing occurs when multiple tasks may access distinct words in the same cache line~\cite{Hoard}, while different tasks may access the same words in true sharing. For both cases, multiple tasks may compete for the shared cache. Cache contention will cause more serious performance degradation, if data has to be loaded from a remote node. 
 
\textbf{Node Imbalance:} When some memory controllers have much more memory accesses than others, it may cause the node imbalance issue.
Therefore, some tasks may wait more time for memory access, thwarting the whole progress of a multithreaded application. 

\textbf{Interconnect Congestion:} Interconnect congestion occurs if some tasks are placed in remote nodes that may use the inter-node interconnection to access their memory. 

\textbf{Remote Accesses:} In a NUMA architecture, local nodes can be accessed with less latency than remote accesses. 
Therefore, it is important to reduce remote access to improve performance.

 
\subsection{Basic Idea}
\label{sec:idea}



Existing NUMA profilers mainly focus on detecting remote accesses, while omitting other performance issues. In contrast, \NP{} has different design goals as follows.  First, it aims to identify different sources of NUMA performance issues, not just limited to remote accesses. Second, \NP{} aims to design architecture- and scheduling-independent approaches that could report performance issues in any NUMA hardware. Third, it aims to provide sufficient information to guide bug fixes.  

For the first goal, \NP{} detects NUMA issues caused by cache contention, node imbalance, interconnect congestion, and remote accesses, where existing work only considers remote accesses.  \textit{Cache contention} can be either caused by false or true sharing, which will impose a larger performance impact and require a different fix strategy. Existing work never separates them from normal remote accesses. In contrast, \NP{} designs a separate mechanism to detect such issues, but tracking possible cache invalidations caused by cache contention. 
It is infeasible to measure all \textit{node imbalance} and \textit{interconnect congestion} without knowing the actual memory and thread binding. Instead, \NP{} focuses on one specific type of issues, which is workload imbalance between different types of threads. 
Existing work omits one type of remote access caused by thread migration, where thread migration will make all local accesses remotely. 
\NP{} identifies whether an application has a higher chance of thread migrations, in addition to normal remote accesses. 
Overall, \NP{} detects more NUMA performance issues than existing NUMA profilers. However, the challenge is to design architecture- and scheduling-independent methods. 

The second goal of \NP{} is to design architecture- and scheduling approaches that do not bind to specific hardware. Detecting remote accesses is based on the key observation of Section~\ref{sec:intro}: if a thread accesses a physical page that was initially accessed by a different thread, then this access will be counted as remote access. This method is not bound to specific hardware, since memory sharing patterns between threads are typically invariant across multiple executions. 
\NP{} tracks every memory access in order to identify the first thread working on each page. Due to this reason, \NP{} employs fine-grained instrumentation, since coarse-grained sampling may miss the access from the first thread. Based on memory accesses, \NP{} also tracks the number of cache invalidations caused by false or true sharing with the following rule: a write on a cache line with multiple copies will invalidate other copies. Since the number of cache invalidations is closely related to the number of concurrent threads, \NP{} divides the score with the number of threads to achieve a similar result with a different number of concurrent threads, as further described in Section~\ref{sec: cacheline}. Load imbalance will be evaluated by the total number of memory accesses of different types of threads. It is important to track all memory accesses including libraries for this purpose. To evaluate the possibility of thread migration, \NP{} proposes to track the number of lock contentions and the number of condition and barrier waits. Similar to false sharing, \NP{} eliminates the effect caused by concurrent threads by dividing with the number of threads. The details of these implementations can be seen in Section~\ref{sec:implementation} . 





For the third goal, \NP{} will utilize the data-centric analysis as existing work~\cite{XuNuma}. That is, it could report the callsite of heap objects that may have NUMA performance issues. 
In addition, \NP{} aims to provide useful information that helps bug fixes, which could be easily achieved when all memory accesses are tracked. \NP{} provides word-based access information for cache contentions, helping programmers to differentiate false or true sharing. It provides threads information on page sharing (help determining whether to use block-wise interleave), and reports whether an object can be duplicated or not by tracking the temporal read/write pattern.
\NP{} also predicts a good thread assignment to achieve better performance for load imbalance issues. In summary, many of these features require fine-grained instrumentation in order to avoid false alarms.

Due to the reasons mentioned above, \NP{} utilizes fine-grained memory accesses to improve the effectiveness and provide better information for bug fixes. \NP{} employs compiler-based instrumentation in order to collect memory accesses due to the performance and flexibility concern. An alternative approach is to employ binary-based dynamic instrumentation~\cite{DynamoRlO, Valgrind, Pin}, which may introduce more performance overhead but without an additional compilation step. \NP{} inserts an explicit function call for each read/write access on global variables and heap objects, while accesses on stack variables are omitted since they typically do not introduce performance issues. To track thread migration, \NP{} also intercepts synchronizations. To support data-centric analysis, \NP{} further intercepts memory allocations to collect their callsites.  
\begin{figure}[!htbp]
\centering
\includegraphics[width=0.98\columnwidth]{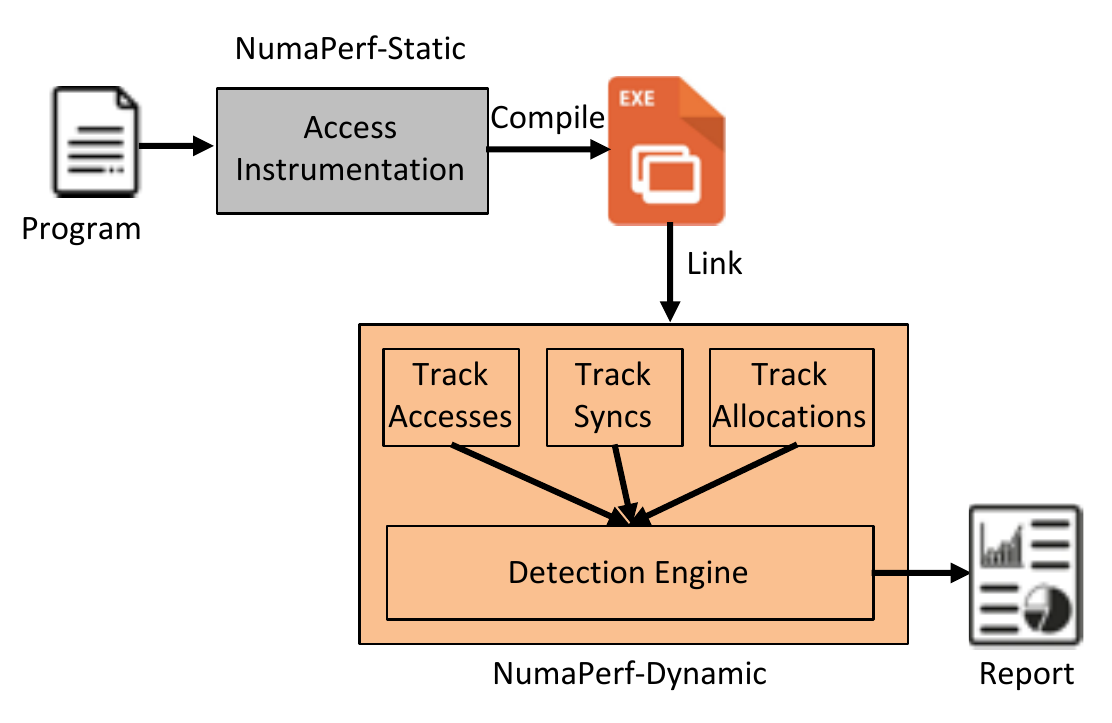}
\caption{Overview of \NP{}\label{fig:overview}}
\vspace{-0.2in}
\end{figure}

Figure~\ref{fig:overview} summarizes \NP{}'s basic idea.  \NP{} includes two components, \NP{}-Static and \NP{}-Dynamic. \NP{}-Static is a static compile-time based tool that inserts a function call before every memory access on heap and global variables, which compiles a program into an instrumented executable file. Then this executable file will be linked to \NP{}-Dynamic so that \NP{} could collect memory accesses, synchronizations, and information of memory allocations. \NP{} then performs detection on NUMA-related performance issues, and reports to users in the end.  More specific implementations are discussed in Section~\ref{sec:implementation}. 

\section{Design and Implementation}
\label{sec:implementation}

This section elaborates \NP{}-Static and \NP{}-Dynamic. \NP{} leverages compiler-based instrumentation (\NP{}-Static) to insert a function call before memory access, which allows \NP{}-Dynamic to collect memory accesses. \NP{} utilizes a pre-load mechanism to intercept synchronizations and memory allocations, without the need of changing programs explicitly. Detailed design and implementation are discussed as follows. 

\subsection{\NP{}-Static} 
\NP{}'s static component (\NP{}-Static) performs the instrumentation on memory accesses. In particular, it utilizes static analysis to identify memory accesses on heap and global variables, while omitting memory accesses on static variables. Based on our understanding, static variables will never cause performance issues, if a thread is not migrated. \NP{}-Static inserts a function call 
upon these memory accesses, where this function is implemented in \NP{}-Dynamic library. In particular, this function provides detailed information on the access, including the address, the type  (i.e., read or write), and the number of bytes.  

\NP{} employs the LLVM compiler to perform the instrumentation~\cite{llvm}. It chooses the intermediate  representation (IR) level for the instrumentation due to the flexibility, since LLVM provides lots of APIs and tools to manipulate the IR. The instrumentation pass is placed at the end of the LLVM optimization passes, where only memory accesses surviving all previous optimization passes will be instrumented.  \NP{}-Static traverses functions one by one, and instruments memory accesses on global and heap variables. The instrumentation is adapted from AddressSanitizer~\cite{AddressSanitizer}.

\subsection{\NP{}-Dynamic}

This subsection starts with tracking application information, such as memory accesses, synchronizations, and memory allocations. Then it discusses the detection of each particular performance issue.  In the following, \NP{} is used to represent \NP{}-Dynamic unless noted otherwise. 

\subsubsection{Tracking Accesses, Synchronizations, and Memory Allocations}

\NP{}-Dynamic implements the inserted functions before memory accesses, allowing it to track memory accesses. Once a memory access is intercepted, \NP{} performs the detection as discussed below. 

\NP{} utilizes a preloading mechanism to intercept synchronizations and memory allocations before invoking correspond functions.
\NP{} intercepts synchronizations in order to detect possible thread migrations, which will be explained later. \NP{} also intercepts memory allocations, so that we could attribute performance issues to different callsites, assisting data-centric analysis~\cite{XuNuma}. For each memory allocation, \NP{} records the allocation callsite and its address range. 
\NP{} also intercepts thread creations in order to set up per-thread data structure. 
In particular, it assigns each thread a thread index.


\subsubsection{Detecting Normal Remote Accesses}


\textit{\NP{} detects a remote access when an access's thread is different from the corresponding page's initial accessor}, as discussed in Section~\ref{sec:overview}. This is based on the assumption that the OS typically allocates a physical page from the node of the first accessor due to the default first-touch policy~\cite{firsttouch}.  Similar to existing work, \NP{} may over-estimate the number of remote accesses, since an access is not a remote one if the corresponding cache is not evicted. 
However, this shortcoming can be overcome easily by only reporting issues larger than a specified threshold, as exemplified in our evaluation (Section~\ref{sec:evaluation}).  

\NP{} is carefully designed to reduce its performance and memory overhead.  \NP{} tracks a page's initial accessor to determine a remote access. A naive design is to employ hash table for tracking such information. 
Instead, \NP{} maps a continuous range of memory with the shadow memory technique~\cite{qinzhao}, which only requires a simple computation to locate the data. 
\NP{} also maintains the number of accesses for each page in the same map. 
We observed that a page without a large number of memory accesses will not cause significant performance issues. Based on this, \NP{} only tracks the detailed accesses for a page, when its number of accesses is larger than a pre-defined (configurable) threshold. 
Since the recording uses the same data structures, \NP{} uses an internal pool to maintain such data structures with the exact size, without resorting to the default allocator.  

For pages with excessive accesses, \NP{} tracks the following information. \textit{First}, it tracks the threads accessing these pages, which helps to determine whether to use block-wise allocations for fixes. 
\textit{Second}, \NP{} further 
divides each page into multiple blocks (e.g., 64 blocks), and tracks the number of accesses on each block.  
This enables us to compute the number of remote accesses of each object more accurately. \textit{Third}, \NP{} further checks whether an object is exclusively read after the first write or not, which could be determined whether duplication is possible or not. 
\textit{Last not least}, \NP{} maintains  word-level information for cache lines with excessive cache invalidations, as further described in Section~\ref{sec: cacheline}.


\textbf{Remote (Access) Score:} \NP{}  proposes a performance metric -- remote  score -- to evaluate the seriousness of remote accesses. An object's remote score is defined as the number of remote accesses within a specific interval, which is currently set as one millisecond. Typically, a higher score indicates more seriousness of remote accesses, as shown in Table~\ref{tab:numa_issues}. 
For pages with both remote accesses and cache invalidations, we will check whether cache invalidation is dominant or not. If the number of cache invalidations is larger than 50\% of remote accesses, then the major performance issue of this page is caused by cache invalidations. We will omit remote accesses instead. 
 


\subsubsection{Detecting False and True Sharing Issues}
\label{sec: cacheline}

Based on our observation,  cache coherence has a higher performance impact than normal remote accesses. Further, false sharing has a different fixing strategy, typically with the padding. 
\NP{} detects false and true sharing separately, which is different from all NUMA profilers. 

\NP{} detects false/true sharing with a similar mechanism as Predator~\cite{Predator}, but adapting it for the NUMA architecture. Predator tracks cache validations as follows: if a thread writes a cache line that is loaded by multiple threads, this write operation introduces a cache invalidation. But this mechanism under-estimates the number of cache invalidations. 
Instead, \NP{} tracks the number of threads  loaded the same cache line, and increases cache invalidations by the number of threads that has loaded this cache line. 

\textbf{False/True Sharing Score:} 
\NP{} further proposes false/true sharing scores for each corresponding object, which is lacked in Predator~\cite{Predator}. The scores are computed by dividing the number of cache invalidations with the product of time (milliseconds) and the number of threads. The number of threads is employed to reduce the impact of parallelization degree, with the architecture-independent method.  
\NP{}  differentiates false sharing from true sharing by recording word-level accesses. 
Note that \NP{} only records word-level accesses for cache lines with the number of writes larger than a pre-defined threshold, due to the performance concern. 

\subsubsection{Detecting Issues Caused by Thread Migration}

As discussed in Section~\ref{sec:intro}, \NP{} identifies applications with excessive thread migrations, which are omitted by all existing NUMA profilers. 
Thread migration may introduce excessive remote accesses. After the migration, a thread is forced to reload all data from the original node, and access its stack remotely afterwards. Further, all deallocations from this thread may be returned to freelists of remote nodes, causing more remote accesses afterwards.    


\textbf{Thread Migration Score:} \NP{} evaluates the seriousness of thread migrations with thread migration scores. This score is computes as the following formula: 
$$S = p \underset{t \in T}{\sum } m_{t} / (rt \cdot \left | T \right |)$$
where $S$ is the thread migration score, $p$ is the parallel phase percentage of the program, $T$ is threads in the program, $\left | T \right |$ is the number of total threads, $m_t$ is the possible migration times for thread $t$, and $rt$ is total running seconds of the program. 

\NP{} utilizes the total number of lock contentions, condition waits, and barrier waits as the possible migration times. The parallel phase percentage indicates the necessarity of performing the optimization. For instance, if the parallel phase percentage is only 1\%, then we could at most improve the performance by 1\%.   In order to reduce the effect of parallelization, the score is further divided by the number of threads. Based on our evaluation, this parameter makes two platforms with different number of threads have very similar results. 

When an application has a large number of thread migrations,  \NP{} suggests users to utilize thread binding to reduce remote accesses. As shown in Table~\ref{tab:numa_issues}, thread migration may degrade the performance of an application (i.e., \texttt{fluidanimate}) by up to 418\%. This shows the importance to eliminate thread migration for such applications.  However, some applications in PARSEC (as not shown in Table~\ref{tab:numa_issues}) have very marginal performance improvement with thread binding. 



\subsubsection{Detecting Load Imbalance}
Load imbalance is another factor that could significantly affect the performance on the NUMA architecture, which could cause node imbalance and interconnect congestion. \NP{} detects load imbalance among different types of threads, which is omitted by existing NUMA-profilers.  


The detection is based on an assumption: \textit{every type of threads should have a similar number of memory accesses in a balanced environment}. \NP{} proposes to utilize the number of memory accesses to predict the workload of each types of threads. In particular, \NP{} monitors memory accesses on heap objects and globals, and then utilizes the sum of such memory accesses to check the imbalance.



\NP{} further predicts an optimal thread assignment with the number of memory accesses. A balance assignment is to balance  memory accesses from each type of threads. For instance, if the number of memory accesses on two type of threads has a one-to-two portion, then \NP{} will suggest to assign threads in one-to-two portion. Section~\ref{sec:casestudies} further evaluates \NP{}'s suggested assignment, where \NP{} significantly outperforms another work~\cite{SyncPerf}.

\section{Experimental Evaluation}
\label{sec:evaluation}

This section aims to answer the following research questions: 

\begin{itemize}
\item \textbf{Effectiveness:} Whether \NP{} could detect more performance issues than existing NUMA-profilers? (Section~\ref{effectiveness}) How helpful of \NP{}'s detection report? (Section~\ref{sec:casestudies})
\item \textbf{Performance:} How much performance overhead is imposed by \NP{}'s detection, comparing to the state-of-the-art tool? (Section~\ref{sec:performance}) 
\item \textbf{Memory Overhead:} What is the memory overhead of \NP{}? (Section~\ref{sec:memory})
\item \textbf{Architecture In-dependence:} Whether \NP{} could detect similar issues when running on a non-NUMA architecture? (Section~\ref{sec:archindependent})	
\end{itemize}


\textbf{Experimental Platform:}  \NP{} was evaluated on a machine with 8 nodes and 128 physical cores in total, except in Section~\ref{sec:archindependent}. This machine is installed with 512GB memory. Any two nodes in this machine are less than or equal to 3 hops, where the latency of two hops and three hops is 2.1 and 3.1 separately, while the local latency is 1.0. The OS for this machine is Linux Debian 10 and the compiler is GCC-8.3.0. The hyperthreading was turned off for the evaluation.

\subsection{Effectiveness}
\label{effectiveness}
\begin{table*}[tp]
 	\setlength{\tabcolsep}{0.45em}
\centering
\begin{tabular}{|c|c|l|l|r|l|l|l|c|c|c|}
    \hline
    \cline{1-9}
    \multirow{2}{*}{Application}& \multirow{2}{*}{Improve}& \multicolumn{7}{c}{Specific Issues}&\\
    \cline{3-10}
    & & \# & Issue & Score & \multicolumn{1}{|c|}{Allocation Site} & \multicolumn{1}{|c|}{Fix Strategy} & \multicolumn{2}{|c|}{Improve} & New \\ 
    \hline 
    
    \multirow{2}{*}{AMG2006}&\multirow{2}{*}{160\%}& 1 & \PS&7390&par\_rap.c:1385&\BI&\multicolumn{2}{|c|}{160\%}& \\
    \cline{3-10}
    
    &  & 2 &\TM&6&&\TB&\multicolumn{2}{|c|}{132\%}&\checkmark \\ \hline

    \multirow{7}{*}{lulesh}&\multirow{7}{*}{594\%}& 3 &\PS&1840&lulesh.cc:543-545&\BI&\multicolumn{2}{|c|}{429\%}& \\
    \cline{3-10}

    &&4&\PS&1504&lulesh.cc:1029-1034&\BI&\multicolumn{2}{|c|}{504\%}&  \\
    \cline{3-10}
    
    &&5&\PS&4496&\multirow{2}{*}{lulesh.cc:2251-2264}&\BI&406\%&\multirow{2}{*}{418\%}& \\
    \cline{3-5}\cline{7-8}\cline{10-10}
    &&6&\FS&26&&\PAD &103\%&& \checkmark\\
    \cline{3-10}
    
    &&7&\PS&1229&\multirow{2}{*}{lulesh.cc:2089}&\BI&392\%&\multirow{2}{*}{407\%}& \\
    \cline{3-5}\cline{7-8}\cline{10-10}
    &&8&\FS&12&&\PAD &104\%&& \checkmark\\
    \cline{3-10}
    
    &&9&\TM&3328&&\TB&\multicolumn{2}{|c|}{382\%}&\checkmark \\ \hline
    
    UMT2013&131\%&10&\TM&18&&\TB&\multicolumn{2}{|c|}{131\%}&\checkmark \\
    \hline 
    \hline
    
    \multirow{3}{*}{bodytrack}&\multirow{3}{*}{109\%}&11&\PS&10800&\multirow{2}{*}{FlexImageStore.h:146}&\PI&\multicolumn{2}{|c|}{\multirow{2}{*}{106\%}}& \\
    \cline{3-5}\cline{7-7}\cline{10-10}
    &&12&\FS&24&& &\multicolumn{2}{|c|}{}&\checkmark \\
    \cline{3-10}
    &&13&\TM&297&&\TB&\multicolumn{2}{|c|}{105\%}&\checkmark \\ \hline
    
    dedup&116\%&14&\TI&&& adjust threads  &\multicolumn{2}{|c|}{116\%}&\checkmark \\ \hline
    
    facesim&105\%&15&\TM&607&&\TB&\multicolumn{2}{|c|}{105\%}&\checkmark \\ \hline
    
    ferret&206\%&16&\TI&&& adjust threads  &\multicolumn{2}{|c|}{206\%}&\checkmark \\ \hline

    \multirow{5}{*}{fluidanimate}&\multirow{5}{*}{429\%}&17&\PS&90534&\multirow{2}{*}{pthreads.cpp:292}& \multirow{2}{*}{\PI} &\multicolumn{2}{|c|}{\multirow{2}{*}{340\%}}& \\
    \cline{3-5} \cline{10-10}
    &&18&\TS&2941&&&\multicolumn{2}{|c|}{}&\checkmark \\
    \cline{3-10}
    
    &&19&\PS&180&\multirow{2}{*}{pthreads.cpp:294}&\PI&112\%&\multirow{2}{*}{160\%}& \\
    \cline{3-5}\cline{7-8}\cline{10-10}
    &&20&\FS&20&&\PAD&158\%&&\checkmark \\
    \cline{3-10}

    &&21&\TM&73&&\TB&\multicolumn{2}{|c|}{418\%}&\checkmark \\ \hline
    
    \multirow{4}{*}{streamcluster}&\multirow{4}{*}{167\%}&22&\PS&427&\multirow{2}{*}{streamcluster.cpp:984}&\PI&100\%&\multirow{2}{*}{103\%}& \\
    \cline{3-5}\cline{7-8}\cline{10-10}
    &&23&\FS&31&&\PAD&102\%&& \checkmark\\
    \cline{3-10}
     
    &&24&\PS&7169&streamcluster.cpp:1845&\DUP&\multicolumn{2}{|c|}{158\%}& \\
    \cline{3-10}
     
    &&25&\TM&229&&\TB&\multicolumn{2}{|c|}{132\%}&\checkmark \\ \hline
    \end{tabular}
  \caption{Detected NUMA performance issues when running on an 8-node NUMA machine. \NP{} detects 15 more performance bugs that cannot be detected by existing NUMA profilers (with a check mark in the last column).}
  \label{tab:numa_issues}
\end{table*}

We evaluated \NP{} on multiple HPC applications (e.g.,  AMG2006~\cite{AMG2006}, lulesh~\cite{LULESH}, and UMT2013~\cite{UMT2013}) and a widely-used multithreaded application benchmark suite --- PARSEC~\cite{parsec}.  Applications with NUMA performance issues are listed in Table~\ref{tab:numa_issues}.  The performance improvement after fixing all issues is listed in ``Improve'' column, with the average of 10 runs, where all specific issues are listed afterwards. For each  issue, the table listed the type of issue and the corresponding score, the allocation site, and the fix strategy. Note that the table only shows cases with page sharing score larger than 1500 (if without cache false/true sharing), false/true sharing score larger than 1, and thread migration score larger than 150. Further, the performance improvement of each specific issue is listed as well. We also present multiple cases studies that show how \NP{}'s report is able to assist bug fixes in Section~\ref{sec:casestudies}.   

Overall, we have the following observations. First, it reports no false positives by only reporting scores larger than a threshold. Second, \NP{} 
detects more performance issues than the combination of all existing NUMA profilers~\cite{Intel:VTune, Memphis, Lachaize:2012:MMP:2342821.2342826, XuNuma, NumaMMA, 7847070, diener2015characterizing, valat:2018:numaprof}. The performance issues that cannot be detected by existing NUMA profilers are highlighted with a check mark in the last column of the table, although some can be detected by specific tools, such as cache false/true sharing issues~\cite{Sheriff, Predator, Cheetah, DBLP:conf/ppopp/ChabbiWL18, helm2019perfmemplus}. This comparison with existing NUMA profilers is based on the methodology. Existing NUMA profilers cannot separate false or true sharing with normal remote accesses, and cannot detect thread migration and load imbalance issues.

When comparing to a specific profiler, \NP{} also has better results even on detecting remote accesses. For lulesh, HPCToolkit detects issues of \# 4 ~\cite{XuNuma}, while \NP{} detects three more issues (\# 3, 5, 7). Fixing these issues improves the performance by up to 504\% (with the threads binding). Multiple reasons may contribute to this big difference. 
 First, \NP{}'s predictive method detects some issues that are not occurred in the current scheduling and the current hardware, while HPCToolkit has no such capabilities. Second, HPCToolkit requires to bind threads to nodes, which may miss remote accesses caused by its specific binding. 
Third, \NP{}'s fine-grained profiling provides a better effectiveness than a coarse-grained profiler like HPCToolkit. 
\NP{} may have false negatives caused by its instrumentation. \NP{} cannot detect an issue of UMT2013 reported by HPCToolkit~\cite{XuNuma}. The basic reason is that \NP{} cannot instrument Fortran code. \NP{}'s limitations are further discussed in Section~\ref{sec:casestudies}.


\subsection{Case Studies}
\label{sec:casestudies}
In this section, multiple case studies are shown how programmers could fix performance issues based on the report. 
\lstset{ %
backgroundcolor=\color{white},      
basicstyle=\footnotesize\ttfamily,  
columns=fullflexible,
tabsize=4,
breaklines=true,               
captionpos=b,                  
commentstyle=\color{mygreen},  
escapeinside={\%*}{*)},        
keywordstyle=\color{blue},     
stringstyle=\color{mymauve}\ttfamily,  
frame=single,
rulesepcolor=\color{red!20!green!20!blue!20},
language=c++,
}

\subsubsection{Remote Accesses}


For remote accesses, \NP{}  not only reports remote access scores, indicating the seriousness of the corresponding issue, but also provides additional information to assist bug fixes. Remote accesses can be fixed with different strategies, such as padding (false sharing), block-wise interleaving, duplication, and page interleaving. 

\begin{lstlisting}[caption={Remote access issue of lulesh },label={blockinterleave},captionpos=b]
Allocation Site: lulesh.cc:2251
Remote score:  4496
False sharing score:  26
True Sharing score:   0.00
Pages accessed by threads:
    0--8, 8--16, 16--23, 23--31 ......
\end{lstlisting}

\NP{} provides a data-centric analysis, as existing work~\cite{XuNuma}. That is, it always attributes performance issues to its allocation callsite. \NP{} also shows the seriousness with its remote access score.

\NP{} further reports more specific information to guide the fix. As shown in  Listing~\ref{blockinterleave}, \NP{} further reports each page that are accessed by which threads. Based on this information,  block-wise interleave is a better strategy for the fix, which achieves a better performance result. However, for Issue 17 or 19 of \texttt{luresh}, there is no such access pattern. Therefore, these bugs can be fixed with the normal page interleave method. 

\begin{lstlisting}[caption={Remote access issue of streamcluster},label={duplicate},captionpos=b]
Allocation site:streamcluster.cpp:1845
Remote score:   7169
False sharing score:  0.00
True Sharing score:   0.00
Continuous reads after the last write:   2443582804
\end{lstlisting}

Listing~\ref{duplicate} shows another example of remote accesses. For this issue (\# 24), a huge number of continuous reads (2330M) were detected after the last write. Based on such a report, the object can be duplicated to different physical nodes, which improves the performance by 158\%, which achieves significantly better performance than page interleave. 

For cache coherency issues,  \NP{} differentiates them from normal remote accesses, and further differentiates false sharing from true sharing. Given the report, programmers could utilize the padding to eliminate false sharing issues. As shown in Table~\ref{sec:evaluation}, many issues have false sharing issues (e.g., \#6, \#8, \#12, \#20, \#23). Fixing them with the padding could easily boost the performance. However, we may simply utilize the page interleave to solve true sharing issues. 


\subsubsection{Thread Migration} 
When an application has frequent thread migrations, it may introduce excessive thread migrations. For such issues, the fix strategy is to bind threads to nodes. Typically, there are two strategies: round robin and packed binding. Round robin is to bind continuous threads to different nodes one by one,  ensuring that different nodes have a similar number of threads. Packed binding is to bind multiple threads to the first node, typically the same as the number of hardware cores in one node, and then to another node afterwards. Based on our observation, round robin typically achieves a better performance than packed binding, which is the default binding policy for our evaluations in Table~\ref{tab:numa_issues}. Thread binding itself achieves the performance improvement by up to 418\% (e.g., \texttt{fluidanimate}), which indicates the importance for some applications. 



\subsubsection{Load Imbalance}

\NP{} not only reports the existence of such issues, but also suggests an assignment based on the number of sampled memory accesses. Programmers could fix them based on the suggestion. 

For \texttt{dedup}, \NP{} reports that memory accesses of anchor, chunk, and compress threads have a proportion of 92.2:0.33:3.43, when all libraries are instrumented. That is, the portion of the chunk and compress threads is around 1 to 10. By checking the code, we understand that \texttt{dedup} has multiple stages, where the anchor is the previous stage of the chunk, and the chunk is the predecessor of the compress. Threads of a previous stage will store results into multiple queues, which will be consumed by threads of its next stage. Based on a common sense that many threads competing for the same queue may actually introduce high contention. Therefore, the fix will simply set the number of chunk threads to be 2. Based on this, we further set the number of compress threads to be 18, and the number of anchor to be 76. The corresponding queues are 18:2:2:4. With this setting, \texttt{dedup}'s performance is improved by 116\%. We further compare its performance with the suggested assignment of another existing work--SyncPerf~\cite{SyncPerf}. SyncPerf assumes that different types of threads should have the same waiting time. SyncPerf proposes the best assignment should be 24:24:48, which could only improve the performance by 105\%. 

In another example of \texttt{ferret}, \NP{} suggests a proportion of $3.3 :1.9 :47.4 :75.3$ for its four types of threads. With this suggestion, we are configuring the threads to be $4 : 2 : 47 : 75$. With this assignment, \texttt{ferret}'s performance increases by 206\% compared with the original version. In contrast, SyncPerf suggests an assignment of $1:1:2:124$
. However, following such an assignment  actually degrades the performance by 354\% instead. 

\subsection{Performance Overhead}
\label{sec:performance}
\begin{figure}[!h]
    \centering
    \includegraphics[width=3.2in]{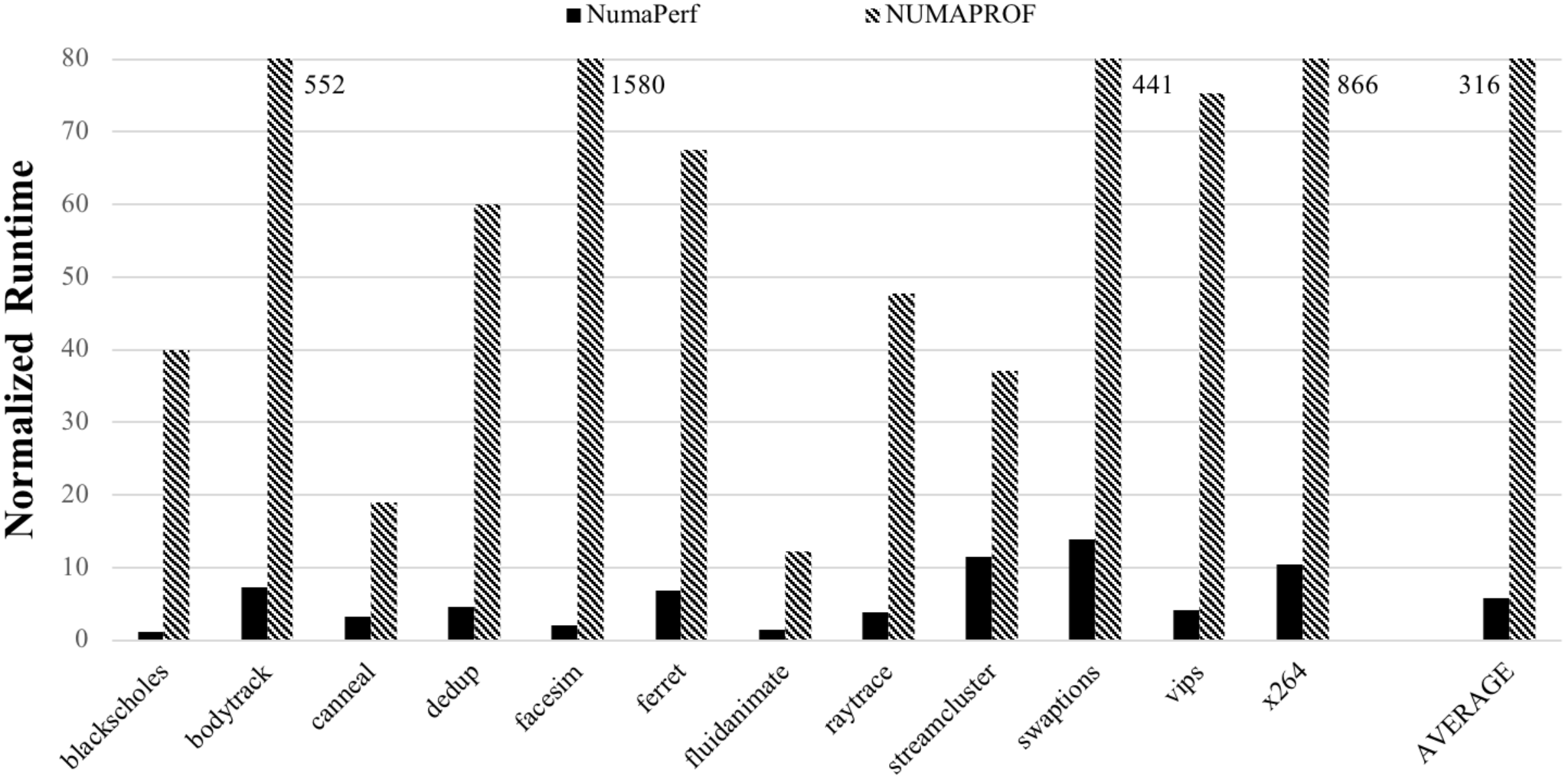}
    \caption{Performance overhead of \NP{} and others.\label{fig:performance}}  
\end{figure}

We also evaluated the performance of \NP{} on PARSEC applications, and the performance results are shown in Figure~\ref{fig:performance}. On average, \NP{}'s overhead is around 585\%, which is orders-of-magnitude smaller than the state-of-the-art fine-grained profiler --- NUMAPROF~\cite{valat:2018:numaprof}. In contrast, NUMAPROF's overhead runs $316\times$ slower than the original one. \NP{} is designed carefully to avoid such high overhead, as discussed in Section~\ref{sec:implementation}. Also, \NP{}'s compiler-instrumentation also helps reduce some overhead by excluding memory accesses on stack variables. 

There are some exceptions. Two applications impose more than $10\times$ overhead, including Swaption and x264. Based on our investigation, the instrumentation with an empty function imposes more than $5\times$ overhead. The reason is that they have significantly more  memory accesses compared with other applications like blackscholes. Based on our investigation, swaption has more than $250\times$ memory accesses than  blackscholes in a time unit. Applications with low overhead can be caused by not instrumenting libraries, which is typically not the source of NUMA performance issues. 


\subsection{Memory Overhead}
\label{sec:memory}
\begin{table}[!htp]
  \centering
    \begin{tabular}{|l|r|r|r|}
    \hline
    \multirow{2}{*}{Apps}&
    \multicolumn{3}{c|}{Memory Usage (MB)}\\
    \cline{2-4}
    &Glibc&\NP&NUMAPROF \\ \hline
    \hline
    blackscholes&617&689&685\\ \hline
    bodytrack&36&139&260\\ \hline
    canneal&887&1476&2383\\ \hline
    dedup&917&1806&2388\\ \hline
    facesim&2638&2826&3005\\ \hline
    ferret&160&301&445\\ \hline
    fluidanimate&470&667&753\\ \hline
    raytrace&1287&1610&2089\\ \hline
    streamcluster&112&216&928\\ \hline
    swaptions&28&67&255\\ \hline
    vips&226&283&463\\ \hline
    x264&2861&3039&3108\\ \hline \hline  
    Total&{\bf 10238}&{\bf 13120}&{\bf 16762}\cr\hline
    \end{tabular}
  \caption{Memory consumption of different profilers. \label{tab:memory_consumption}}
  \vspace{-0.2in}
\end{table}

We further evaluated \NP{}'s memory overhead with PARSEC applications. The results are shown in Table~\ref{tab:memory_consumption}. In total, \NP{}'s memory overhead is around 28\%, which is much smaller than the state-of-the-art fine-grained profiler --- NUMAPROF~\cite{valat:2018:numaprof}. \NP{}'s memory overhead is mainly coming from the following resources. First, \NP{} records the detailed information in page-level and cache-level, so that we could provide detailed information to assist bug fixes. Second, \NP{} also stores allocation callsites for every object in order to attribute performance issues back to the data. 

We  notice that some applications have a larger percentage of memory overhead, such as \texttt{streamcluster}. For it, a large object has very serious NUMA issues. Therefore, recording page and cache level detailed information contributes to the major memory overhead. However, overall, \NP{}'s memory overhead is totally acceptable, since it provides much more helpful information to assist bug fixes. 



\subsection{Architecture Sensitiveness}
\label{sec:archindependent}

We further confirm whether \NP{} is able to detect similar performance issues when running on a non-NUMA or UMA machine. We further performed the experiments on a two-processor machine, where each processor is Intel(R) Xeon(R) Gold 6230 and each processor has 20 cores. We explicitly disabled all cores in node 1 but only utilizing 16 hardware cores in node 0. This machine has 256GB of main memory, 64KB L1 cache, and 1MB of L2 cache. The experimental results are further listed in Table~\ref{tab:independent}. For simplicity, we only listed the applications, the issue number, and serious scores in two different machines. 

Table~\ref{tab:independent} shows that most reported scores in two machines are very similar, although with small variance. The small variance could be caused by multiple factors, such as parallelization degree (concurrency). However, this table shows that all serious issues can be detected on both machines. This indicates that \NP{}  achieves its design goal, which could even detect NUMA issues without running on a NUMA machine.

\begin{table}[!htp]
 	\setlength{\tabcolsep}{0.45em}
\centering
\begin{tabular}{|c|l|l|l|l|}
\hline
\multirow{2}{*}{Application}  & \multicolumn{4}{c|}{Specific   Issues}                                                              \\ \cline{2-5} 
                              & \# & \multicolumn{1}{c|}{Type} & \multicolumn{1}{c|}{\specialcell{Score\\(NUMA)}} & \multicolumn{1}{c|}{\specialcell{Score\\(UMA)}} \\ \hline
\multirow{2}{*}{AMG2006}       & 1  & \PS     & 7390  & 5405  \\ \cline{2-5} 
                               & 2  & thread migration & 6     & 6      \\ \hline
\multirow{7}{*}{lulesh}        & 3  & \PS     & 1840  & 2443  \\ \cline{2-5} 
                               & 4  & \PS     & 1504  & 2353  \\ \cline{2-5} 
                               & 5  & \PS     & 4496  & 4326  \\ \cline{2-5} 
                               & 6  & false sharing    & 26    & 51   \\ \cline{2-5} 
                               & 7  & \PS     & 1229  & 2136  \\ \cline{2-5} 
                               & 8  & false sharing    & 12    & 27    \\ \cline{2-5} 
                               & 9  & thread migration & 3328  & 5213      \\ \hline
UMT2013                        & 10 & thread migration & 18    &      \\ \hline \hline
\multirow{3}{*}{bodytrack}     & 11 & \PS     & 10800 & 8203  \\ \cline{2-5} 
                               & 12 & false sharing    & 24    & 153   \\ \cline{2-5} 
                               & 13 & thread migration & 297   & 190   \\ \hline 
dedup                          & 14 & thread imbalance &   92:1:3    &  88:4:4     \\ \hline
facesim                        & 15 & thread migration & 607   & 274   \\ \hline
ferret*                          & 16 & thread imbalance &      &       \\ \hline
\multirow{5}{*}{fluidanimate} & 17 & \PS              & 90534                            & 15765 \\ \cline{2-5} 
                               & 18 & true sharing     & 2941  & 1753  \\ \cline{2-5} 
                               & 19 & \PS     & 180   & 95   \\ \cline{2-5} 
                               & 20 & false sharing    & 20    & 80    \\ \cline{2-5} 
                               & 21 & thread migration & 73    & 34    \\ \hline
\multirow{4}{*}{streamcluster} & 22 & \PS     & 427   & 270   \\ \cline{2-5} 
                               & 23 & false sharing    & 31    & 153   \\ \cline{2-5} 
                               & 24 & \PS     & 7169  & 10259 \\ \cline{2-5} 
                               & 25 & thread migration & 229   & 214   \\ \hline
\end{tabular}
  \caption{Evaluation on architecture Sensitiveness. We evaluated \NP{} on a non-NUMA (UMA) machine, which has very similar results as that on a NUMA machine. For \texttt{ferret}, \NP{} reports a proportion of  $3:2:48:75$ on the 8-node NUMA machine, and $5:4:50:77$ on the UMA machine. \label{tab:independent}}
  \vspace{-0.2in}
\end{table}

\section{Limitation}
\label{sec:discussion}
\NP{} bases on compiler-based instrumentation to capture memory accesses. Therefore, it shares the same shortcomings and strengths of all compiler-based instrumentation. On the one side, \NP{} can perform static analysis to reduce unnecessary memory accesses, such as accesses of stack variables. \NP{} typically achieves much better performance than binary-based instrumentation tools, such as Numaprof~\cite{valat:2018:numaprof}. On the other side, \NP{} requires the re-compilation (and the availability of the source code), and will miss memory accesses without the instrumentation. That is, it can not detect NUMA issues caused by non-instrumented components (e.g., libraries), suffering from false negatives. However, most issues should only occur in applications, but not libraries. 
\section{Related Work}
\label{sec:related}

This section discusses NUMA-profiling tools at first, and then discusses other relevant tools and systems.
\subsection{NUMA Profiling Tools}


\paragraph{Simulation-Based Approaches:}
 Bolosky et al. propose to model NUMA performance issues based on the collected trace, and then derive a better NUMA placement policy~\cite{Bolosky:1991:NPR:106972.106994}.
 NUMAgrind employs binary instrumentation to collect memory traces, and simulates cache activities and page affinity~\cite{NUMAGrind}. MACPO reduces the overhead of collecting memory traces and analysis by focusing on code segments that have known performance bottlenecks~\cite{MACPO}. That is, it typically requires programmer inputs to reduce its overhead.  Simulation-based approaches could be utilized for any architecture, which are very useful. However, they are typically extremely slow, with thousands of performance slowdown, which makes them un-affordable even for development phases. Further, they still require to evaluate the performance impact for a given architecture, which will significantly limit its usage. \NP{} utilizes a measurement based approach, which avoids significant performance overhead of simulation-based approaches. 

\paragraph{Fine-Grained Approaches:} 
TABARNAC focuses on the visualization of memory access behaviors of different data structures~\cite{TABARNAC}. It uses PIN to collect memory accesses of every thread on the page level, and then relates with data structure information together to visualize the usage of data structures. It introduces the runtime overhead between $10\times$ and $60\times$, in addition to its offline overhead. Diener et al. propose to instrument memory accesses with PIN dynamically, and then characterize distribution of accesses of different NUMA nodes~\cite{diener2015characterizing}. The paper does not present the detailed overhead. 
Numaprof also uses the binary instrumentation  (i.e., PIN) to collect and identify local and remote memory accesses~\cite{valat:2018:numaprof}. 
Numaprof relies on a specific thread binding to detect remote accesses, which shares the same shortcoming as other existing work~\cite{XuNuma, 7847070}. 
Numaprof also shares the same issues with other tools, which only focuses on remote accesses while omitting other issues such as cache coherence issues and imbalance issues. 
In addition, Numaprof is only a code-based profiler that could only report program statements with excessive remote memory access, which requires programmers to figure out the data (object) and a specific strategy. Due to this shortcoming, it makes the comparison with Numaprof extremely difficult and time-consuming. 
In contrast, although \NP{} also utilizes fine-grained measurement, it detects more issues that may cause performance issues in any NUMA architecture, and provides more useful information for bug fixes.    


\paragraph{Coarse-Grained Approaches:}
Many tools employ hardware Performance Monitoring Units (PMU) to identify NUMA-related performance issues, such as VTune~\cite{Intel:VTune}, Memphis~\cite{Memphis}, MemProf~\cite{Lachaize:2012:MMP:2342821.2342826}, Xu et al.~\cite{XuNuma}, NumaMMA~\cite{NumaMMA}, and LaProf~\cite{7847070}, where their difference are further described in the following. 
Both VTune~\cite{Intel:VTune} and Memphis~\cite{Memphis} only detects NUMA-performance issues on statically-linked variables.  
MemProf proposes the employment of hardware Performance Monitoring Units (PMU) to identify NUMA-related performance issues~\cite{Lachaize:2012:MMP:2342821.2342826}, with the focus on remote accesses. It constructs data flow between threads and objects to help understand NUMA performance issues. One drawback of MemProf is that it requires an additional kernel module that may prevent people of using it. Similarly, Xu et al. also employ PMU to detect NUMA performance issues~\cite{XuNuma}, but without the change of the kernel. It further proposes a new metric, the NUMA latency per instruction, to evaluate the seriousness of NUMA issues. This tool has a drawback that it statically binds every thread to each node, which may miss some NUMA issues due to its static binding. 
NumaMMA also collects traces with PMU hardware, but focuses on the visualization of memory accesses ~\cite{NumaMMA}. LaProf focuses on multiple issues that may cause performances issues in NUMA architecture~\cite{7847070}, including data sharing, shared resource contention, and remote imbalance. LaProf has the same shortcoming by binding every thread statically.  Overall, these sampling-based approaches although imposes much lower overhead, making them applicable even for the production environment, they cannot detect all NUMA performance issues especially when most of them only focus on remote accesses. In contrast, \NP{} aims to detect performance issues inside  development phases, avoiding any additional runtime overhead. Also, \NP{} focuses more aspects with a predictive approach, not just limited to remote accesses in the current hardware. Our evaluation results confirm \NP{}'s comprehensiveness and effectiveness. 


\subsection{Other Related Tools}
RTHMS also employs PIN to collect memory traces, and then assigns a score to each object-to-memory based on its algorithms~\cite{RTHMS}. It aims for identifying the peformance issues for the hybrid DRAM-HBM architecture, but not the NUMA architecture, and has a higher overhead than \NP{}. Some tools focus on the detection of false/true sharing issues~\cite{Sheriff, Predator, Cheetah, DBLP:conf/ppopp/ChabbiWL18, helm2019perfmemplus}, but skipping other NUMA issues. 

SyncPerf also detects load imablance and predicts the optimal thread assignment~\cite{SyncPerf}. SyncPerf aims to achieve the optimal thread assignment by balancing the waiting time of each types of threads. In contrast, \NP{} suggests the optimal thread assignment based  the number of accesses of each thread, which indicates the actual workload. 



\section{Conclusion}
\label{sec:conclusion}

Parallel applications running on NUMA machines are prone to different types of performance issues. Existing NUMA profilers may miss significant portion of optimization opportunities. Further, they are bound to a specific NUMA topology. Different from them, \NP{} proposes an architecture-independent  and scheduling-independent method that could detect NUMA issues even without running on a NUMA machine. Comparing to existing NUMA profilers, \NP{} detects more performance issues without false alarms, and also provides more helpful information to assist bug fixes. In summary, \NP{} will be an indispensable tool that could identify NUMA issues in development phases.

\bibliographystyle{plain}
\bibliography{refs,tongping}

\end{document}